\documentclass[twocolumn]{aastex631}

\newcommand{\angstrom}{\text{\normalfont\AA}}

\begin{document}

\title{The Origin of the Consistent Planetary Nebula Luminosity Function Bright-end Cutoff}

\author[0000-0003-3024-7218]{Philippe Z. Yao}\thanks{philippe.yao@princeton.edu}
\affiliation{Department of Astrophysical Sciences, Princeton University, Peyton Hall, Princeton, NJ 08544, USA}

\author[0000-0001-9185-5044]{Eliot Quataert}
\affiliation{Department of Astrophysical Sciences, Princeton University, Peyton Hall, Princeton, NJ 08544, USA}



\begin{abstract}

The [O III] 5007 $\angstrom$ line is typically the brightest line in planetary nebula (PN) spectra.  Observations show that the brightest [O III] 5007 $\angstrom$ PN in a galaxy -- the planetary nebula luminosity function (PNLF) bright-end cutoff -- is surprisingly independent of galaxy type. To understand the origin of this puzzling uniformity, we simulate PNe with a range of cloud and star parameters using the photoionization code CLOUDY. We find that the peak [O III] 5007 $\angstrom$ luminosity depends weakly on both the central stellar effective temperature at high temperature and on the total PN ejecta mass; however, the peak [O III] 5007 $\angstrom$ luminosity depends strongly on the central stellar luminosity and the PN dust-to-gas mass ratio.  We explain these scalings physically.   They imply that a higher dust-to-gas mass ratio at higher central stellar luminosity can help explain a constant bright-end cutoff in the PNLF across galaxy types. This prediction is testable with a survey of galactic PNe. The surviving remnants of double white dwarf mergers should also produce photoionized nebulae analogous to PNe. These may be preferentially present at the high luminosity end of the [O III] PLNF and could explain the existence of PNe in early-type galaxies that are more luminous in [O III] than expected from single-star evolutionary models. The presence of white dwarf mergers in both young and old stellar populations could contribute to the uniformity of the [O III] PNLF across galaxy types; such nebulae would lack the hydrogen lines otherwise characteristic of PNe.

\end{abstract}

\keywords{Planetary nebulae (1249) --- White dwarf stars (1799) ---- Stellar mergers (2157) }


\section{Introduction}
Planetary nebulae (PNe) are signposts of white dwarf (WD) formation known for their unique photoionized spectra \citep[see review by][and references therein]{Peimbert2017}. The [O III] 5007 $\angstrom$ line in a PN can reprocess  $\sim 10-13\%$ of the total luminosity of the central star \citep[e.g.][]{Dopita1992,Ciardullo2010,Sch2010}. This feature enables extragalactic surveys of PNe populations in nearby galaxies and galaxy clusters. Not only are PNe very useful as tracers of galaxy stellar populations \citep[e.g.][]{Buzzoni2006,Douglas2007,Hartke2020} and kinematics \citep[e.g.][]{Aniyan2018,Aniyan2021}, but their luminosity function can be used as a distance indicator to nearby galaxies \citep[e.g.][]{Jacoby1989,Ciardullo1989,Jacoby1992,Roth2021,Spriggs2021}. This is due to the invariance of the bright-end cutoff of the planetary nebula luminosity function (PNLF) at an absolute magnitude $M_{5007} \simeq -4.5 \, \rm mag$. This peak [O III] 5007 $\angstrom$ luminosity for the PNe population is roughly constant across  galaxy types \citep{deGrijs2013}. 

There is no consensus regarding the physical origin of the rough constancy of the bright-end cutoff to the PNLF (e.g., \citealt{Ciardullo2012,Davis2018,Gesicki2018,Valenzuela2019,Kwitter2022,Souropanis2023}).  \citet{Ciardullo2012} outlined a number of theoretical reasons that work against a consistent cutoff luminosity; most notably, early-type galaxies do not host massive stars like star-forming late-type galaxies.   The resulting newly formed white dwarfs are thus fainter in early-type galaxies than in late-type galaxies, naively leading to a significant difference between the maximum brightness of the reprocessed [O III] 5007 $\angstrom$ line. This drastic difference is not, however, observed. One factor that decreases the contribution of more massive stars to the PNLF is that the lifetime of massive post-AGB stars is significantly shorter than that of lower mass post-AGB stars \citep{Jacoby1989,Gesicki2018}.  This alone, however, is insufficient to explain the observed constancy of the bright-end cutoff to the PNLF. It is possible that updated stellar evolution models \citep[such as in][]{Miller2016} could resolve this issue, though this also depends on the exact time evolution of the surrounding nebula as the star evolves from the AGB to the WD cooling sequence and the mass fraction of $3-6$ Gyr-old stars in elliptical galaxies \citep{Gesicki2018}.  \citet{Davis2018} suggested that dust could be important for the PNLF bright-end cutoff due to potentially different amounts of mass loss for AGB stars with different progenitor masses. In this paper, we further quantify the effect of dust on the PNLF.

Binary evolution has also often been invoked to explain the constant PNLF bright-end cutoff, in the form of blue stragglers, common envelope interactions, symbiotic stars, and accreting WDs \citep[e.g.][respectively]{Ciardullo2005, Soker2006, Davis2018, Souropanis2023}.  One source of cooling WDs and photoionized nebulae not typically accounted for in models of the PNLF is WD merger remnants.   The fraction of double WD mergers that survive the merger without producing a Type Ia supernovae is uncertain.  However, if the merger of two Carbon-Oxygen (CO) WDs or a CO and ONeMg WD does not produce an explosion, the merger remnant evolves to become a hot luminous proto-WD, likely with surrounding photoionized nebula analogous to a PN but lacking strong H lines \citep{Schwab2016,Schwab2021,Yao2023}.
Since WD mergers occur in both young and old stellar populations, the nebulae produced by surviving WD merger remnants may be more uniform across galaxy types than PNe produced by single stellar evolution.

Here, we use the CLOUDY spectral synthesis code\footnote{\url{https://gitlab.nublado.org/cloudy/cloudy}} \citep[v17.02; ][]{CLOUDY1998,Ferland2017} to perform a parameter-space study of the dependence of the [O III] 5007 $\angstrom$ luminosity ($L_{5007}$) in PNe.  We also evaluate the possible contribution of surviving WD merger remnants to the PN luminosity function.  We first summarize our grid of and results from CLOUDY simulations in \S \ref{sec:sims}. We then model the $L_{5007}$ dependence and outline the physical reasons behind these correlations in \S \ref{sec:physics}. We consider WD merger remnant nebulae in \S \ref{sec:merger}.  And finally, we summarize this article and discuss observational tests in \S \ref{sec:discussion}.

\section{Photoionization Calculations} \label{sec:sims}

    \begin{figure}
        \centering
        \includegraphics[width=\columnwidth]{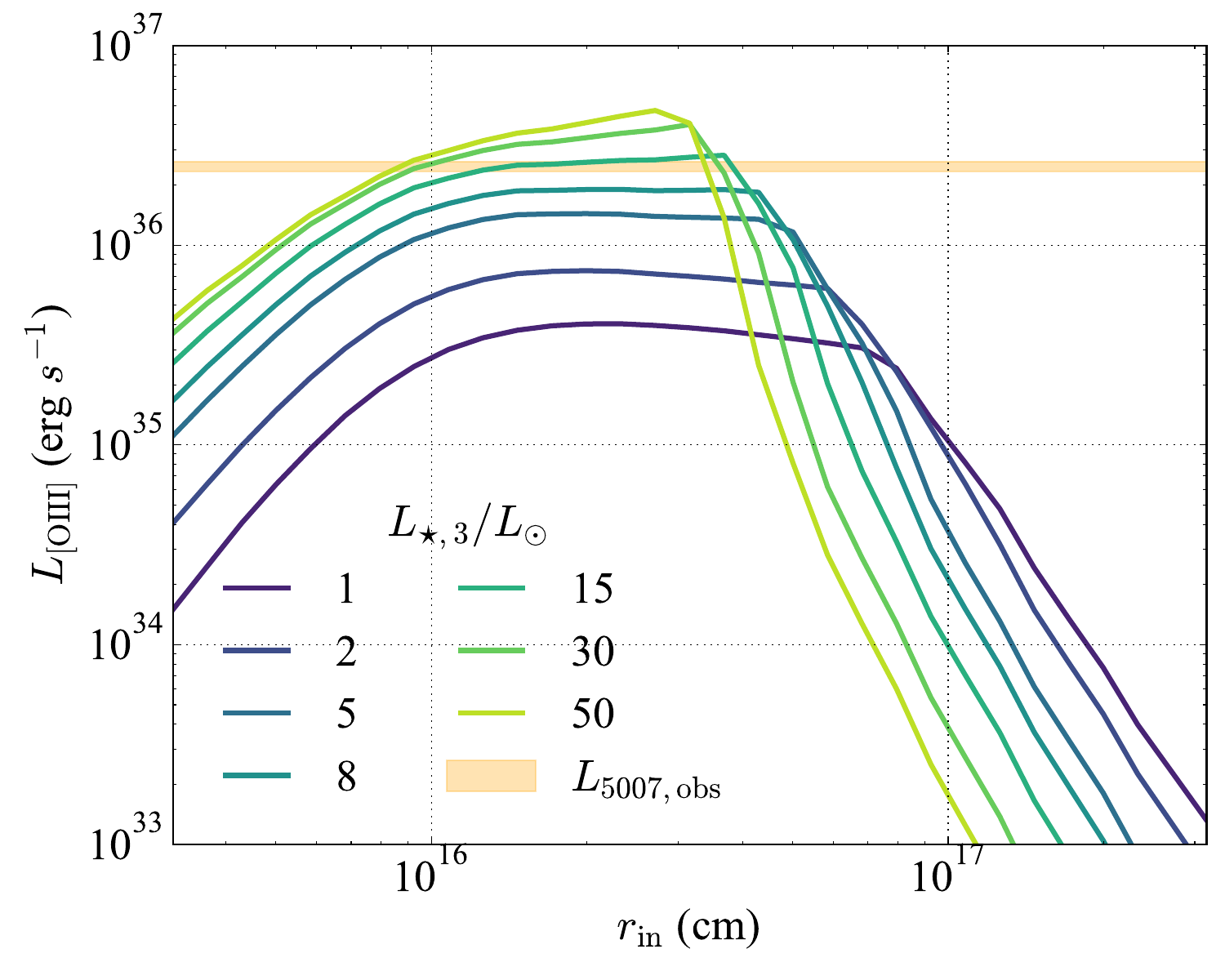}
        \caption{$L_{\rm [O\,III]}$ from CLOUDY simulations as a function of cloud inner radii at a range of central stellar luminosity. The D/G mass ratio is assumed to be 1\% in this case. The peak line luminosity occurs at similar radii, and is used to produce Figure \ref{fig:PNLF} at different ejecta masses. The orange shaded region reflects the observed PNLF bright-end cutoff at $M_{5007} = -4.53\pm0.06$ translated to line luminosity from \citet{Ciardullo2012}.}
        \label{fig:radial}
    \end{figure}

At the end stage of low-mass stellar evolution, the outer layers of a star expand into a PNe, at the center of which is a luminous and increasingly hotter post-AGB star. One of the resultant emission lines from the surrounding photoionized nebula, the [O III] 5007 $\angstrom$ line, is the most important coolant for PNe. Here we assess what physical parameters the [O III] line luminosity depends most sensitively on. We model the photoionized nebula with CLOUDY, assuming the central star is a black body, and the nebula is a spherical shell of ejecta with a thickness $\sim 3 \times$ the inner radius ($r_{\rm in}$). We also assume solar composition and uniform density throughout the cloud. To study which parameters of the PNe affect the bright-end cutoff the most, we vary the central stellar luminosity ($L_{\star}$), stellar effective temperature ($T_{\rm eff}$), ejecta mass ($m_{\rm ej}$), and dust-to-gas mass ratio ($R_{\rm D/G}$); then we find where the [O III] 5007 $\angstrom$ luminosity peaks as a function of the size of the nebula $r_{\rm in}$, given a set of parameters. An example is provided in Figure \ref{fig:radial}, where models typically peak in $L_{\rm [O\,III]}$ between $r_{\rm in} \sim 10^{16}-10^{17} \rm cm$.

\begin{figure*}
        \centering
        \includegraphics[width=\textwidth]{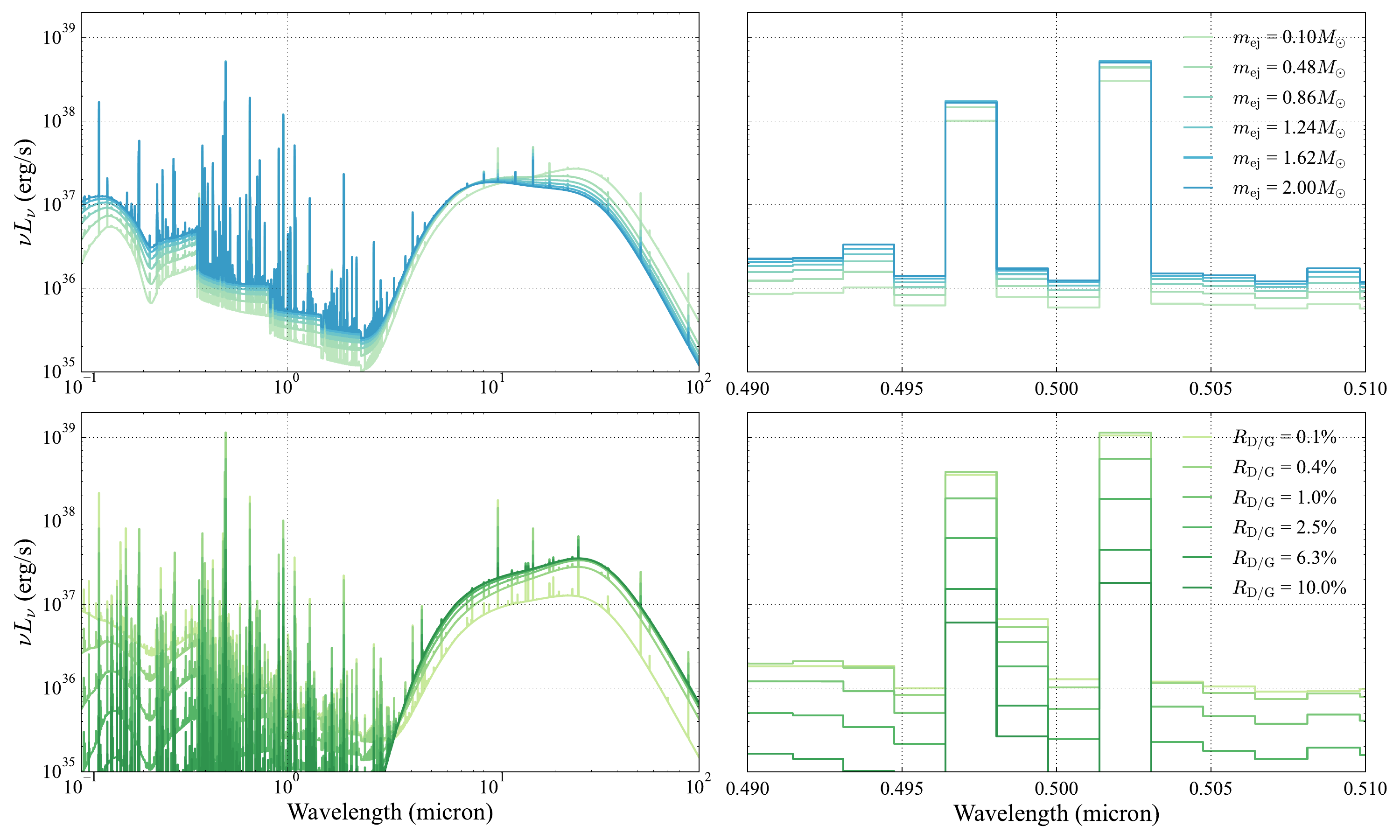}
        \caption{CLOUDY spectra for PNe modeled with a range of ejecta mass at constant D/G mass ratio (top) and with a range of D/G mass ratio at constant ejecta mass (bottom). For comparison, we assume the same luminosity of $13,000 L_{\odot}$ and effective temperature of $T_{\rm eff} = 10^5 \rm K$ for the central star. The left panels cover a wider spectral range from $0.1 \mu \rm m$ to 100 $\mu \rm m$, while the right panels are centered on the [O III] 4959, 5007 $\angstrom \angstrom$ doublet.}
        \label{fig:spec}
    \end{figure*}

Oxygen-rich AGB stars produce PNe with oxygen-rich grains such as silicates and carbon-rich AGB stars produce carbon-rich grains such as graphite \citep{Peimbert2017}. Since dust is expected to play an important role in PNe evolution and their observational signatures \citep{SS1999}, we show simulations that include graphite grains resolved into ten size bins by default following \citet{Mathis1977}'s power-law distribution. We note that while grains are crucial to reproducing the bright-end cutoff in our studies, the choice of grain species does not significantly alter the [O III] 5007 $\angstrom$ luminosity.   We show the effect of dust with CLOUDY spectra in Figure \ref{fig:spec} when varying $R_{\rm D/G}$ or $m_{\rm ej}$, while keeping the other parameter fixed. On the left panel, reprocessed dust IR emission clearly dominates at longer wavelengths; and in the right panel, we zoom into the [O III] 4959, 5007 $ \angstrom \angstrom$ doublet to show the change in [O III] line luminosity as a function of $R_{\rm D/G}$ and $m_{\rm ej}$.  Note that there is little dependence of the [O III] line luminosity on ejecta mass but a strong dependence on the dust-to-gas mass ratio; we explain this physically below.      In our calculations, we assume dust and gas are perfectly mixed, but in reality viewing angle and inhomogeneous dust distributions can have an effect on the emission line strengths. For reasons explained below it is unlikely that this will change our core result that the [O III] line luminosity depends primarily on the dust-to-gas mass ratio $R_{\rm D/G}$ rather than the absolute dust or ejecta mass.

\section{Reproducing the bright-end cutoff}\label{sec:physics}

    \begin{figure*}
        \centering
             \includegraphics[width=\textwidth]{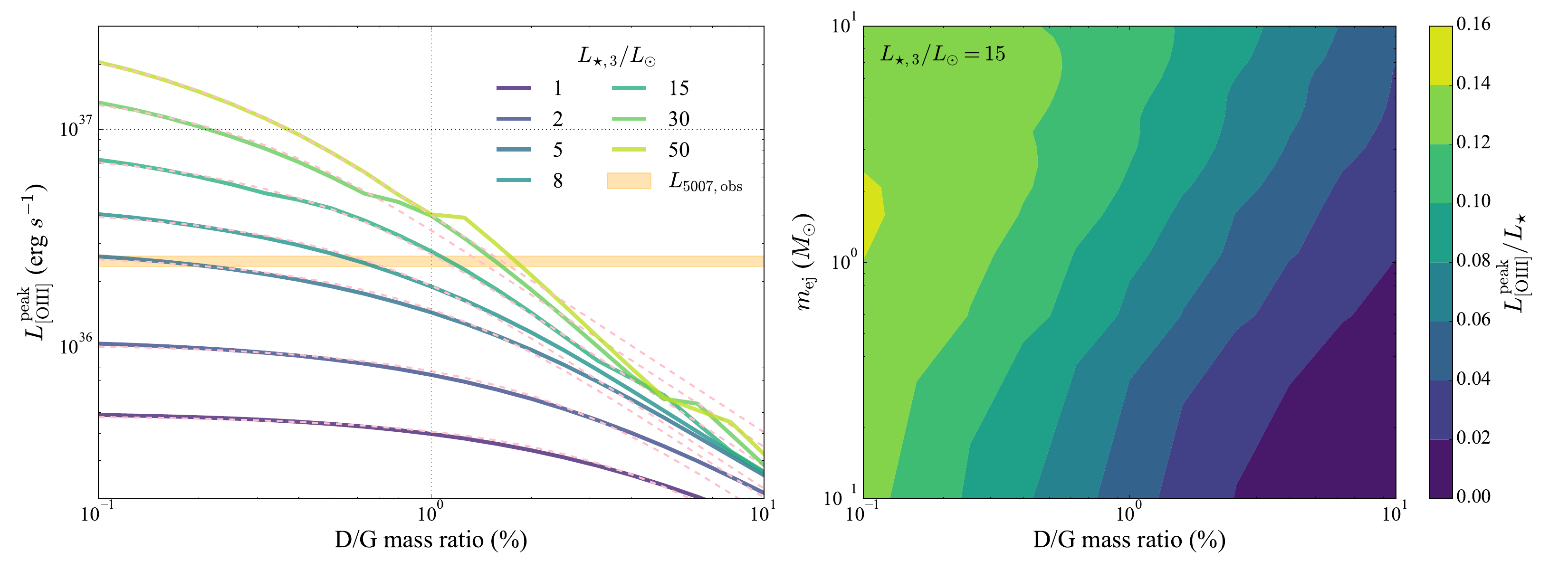}
        \caption{Peak [O III] 5007 $\angstrom$ luminosity ($L^{\rm peak}_{\rm [O\,III]}$) produced by a standard solar metallicity PNe simulated with CLOUDY. The stellar luminosity $L_{\star,3}$ in the legend is in units of $10^3 L_{\odot}$. The left panel assumes an ejecta mass of $m_{\rm ej} = 0.1 M_{\odot}$ and an effective stellar temperature $T_{\rm eff} = 10^5 \rm K$, and varies the dust-to-gas mass ratio and central stellar luminosity. The orange-shaded region reflects the observed PNLF bright-end cutoff at $M_{5007} = -4.53\pm0.06$ translated to line luminosity from \citet{Ciardullo2012}. The right panel plots $L^{\rm peak}_{\rm [O\,III]}$ as its ratio to the center stellar luminosity $L_{\star}$. Here we assume a central stellar luminosity of $15,000 L_{\odot}$ and an effective stellar temperature $T_{\rm eff} = 10^5 \rm K$, and vary the ejecta mass and the dust-to-gas mass ratio.}
        \label{fig:PNLF}
    \end{figure*}

\subsection{Simulation Results}

Figure \ref{fig:PNLF} shows our results for the dependence of the peak [O III] line luminosity $L^{\rm peak}_{\rm [O\,III]}$ on stellar luminosity, ejecta mass, and dust to gas mass ratio.  We find little dependence on 
$T_{\rm eff}$ once the central star is hotter than $\sim 10^5 K$ and so do not plot models with different $T_{\rm eff}$.   The left panel of Figure \ref{fig:PNLF} shows models with $m_{\rm ej} = 0.1 M_{\odot}$ with a range of $L_{\star}$ from $1000 L_{\odot}$ to $50,000 L_{\odot}$ and $R_{\rm D/G}$ from 0.1\% to 10\%. On the right panel of Figure \ref{fig:PNLF}, we show a grid of calculations for $L_{\star} = 15,000 L_{\odot}$ while varying $R_{\rm D/G}$ and $m_{\rm ej}$. Each value of $L^{\rm peak}_{\rm [O\,III]}$ is found by calculating a grid of models for varying nebula sizes as in Figure \ref{fig:radial}. Figure \ref{fig:PNLF} shows a strong dependence of the peak [O III] line luminosity on both the central stellar luminosity $L_{\star}$ and dust to gas mass ratio $R_{\rm D/G}$ but a comparatively weak dependence on ejecta mass $m_{\rm ej}$. The peak [O III] line luminosity is typically $\sim 10\%$ of the total stellar luminosity, similar to the observed maximum ratios ($L_{\rm [O\,III]}/L_{\star}$) in \citet{Ciardullo2010} and \citet{Sch2010}.  The exception to this is at high D/G mass ratios $R_{\rm D/G} \gtrsim 1 \%$ where dust suppresses $L_{\rm [O\,III]}/L_{\star}$.   

We find that the following functional form provides a reasonable fit to the simulation results:
\begin{equation}
    \label{eq:LOIII}
L^{\rm peak}_{\rm [O\,III]} \propto L_\star \frac{\left(1-e^{-\tau_{\rm dust}}\right)}{\tau_{\rm dust}}
\end{equation}
where
\begin{equation}
    \label{eq:taudust}
    \tau_{\rm dust} = b R_{\rm D/G} L_{\star}^{2/3} m_{\rm ej}^{-1/3}
\end{equation}
and $b$ is a constant that varies weakly with other parameters.   The fits are plotted with pink dotted lines in the left panel of Figure \ref{fig:PNLF}.

Equation \ref{eq:LOIII} is the solution of the radiation transfer equation for a uniform emissivity in the [O III] line attenuated internally by dust with optical depth $\tau_{\rm dust}$ given by equation \ref{eq:taudust}.    
$L^{\rm peak}_{\rm [O\,III]} \propto L_\star$ because the latter sets the total heating rate of the PNe by photoionization.  $L^{\rm peak}_{\rm [O\,III]}$ is independent of $T_{\rm eff}$ for $T_{\rm eff} \gtrsim 10^5$ K because at such high temperatures (and at the small PNe sizes that maximize $L^{\rm peak}_{\rm [O\,III]}$) almost all of the incident radiation is reprocessed by photoionization, independent of $T_{\rm eff}$.    The weak dependence of the peak [O III] luminosity on the ejecta mass arises as follows.   The peak [O III] luminosity in Figure \ref{fig:radial} occurs for nebula sizes comparable to the Stromgren sphere radius $R_S$.   Roughly independent of $m_{\rm ej}$ a fixed fraction of the central stellar luminosity is radiated in the [O III] line, though the nebula size at which this peak luminosity is realized varies $\propto R_S \propto m_{\rm ej}^{2/3} L^{-1/3}$. 
Because the peak [O III] luminosity occurs when the nebula size is $\sim R_S$, the dust optical depth in equation \ref{eq:taudust} is evaluated at this radius.     When the PN has a size $\gtrsim R_S$ the nebula can be photoionized by just a small fraction of the central star's radiation and so most of the ionizing photons escape rather than being reprocessed into optical line emission (a `matter bounded' nebula).   This is why $L_{\rm [O\,III]}$ declines rapidly at large $r_{\rm in}$ in Figure \ref{fig:radial}. The exact functional dependence on $\tau_{\rm dust}$ in equation \ref{eq:LOIII} is specific to well-mixed dust and gas, and is likely more complicated in reality. However, it is also likely that a key qualitative conclusion of equation \ref{eq:LOIII} and Figure \ref{fig:spec} will hold more broadly, namely that the strongest dependence of the [O III] line luminosity is on dust-to-gas mass ratio rather than total ejecta (or dust) mass. This is because to first approximation the ejecta mass only sets the physical size of the nebula at the time it transitions from ionization to matter-bounded, not the total photoionized luminosity of the nebula.

\subsection{PNLF Bright-end cutoff}

    \begin{figure}
        \centering
        \includegraphics[width=\columnwidth]{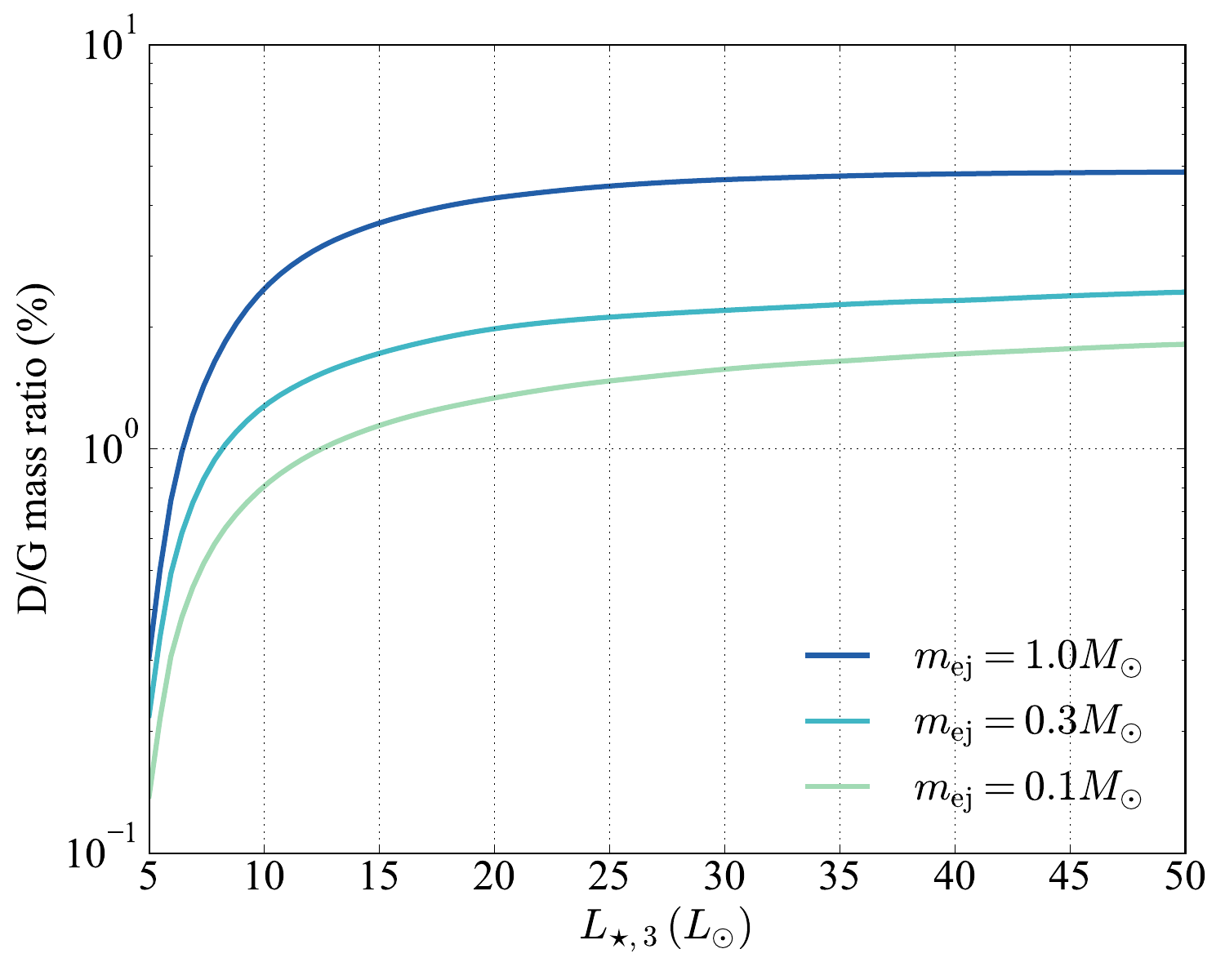}
        \caption{D/G mass ratio ($R_{\rm D/G}$) needed for a range of central stellar luminosities to produce $L^{\rm peak}_{\rm [O\,III]}$ that matches the observed PNLF cutoff magnitude, for models with $m_{\rm ej} = 0.1,\,0.3,\, \& 1.0 M_{\odot}$ For reference, $1-3 M_{\odot}$ progenitors can produce PNe hosting post-AGB stars of $\sim 8\times 10^3 L_{\odot}$, and $\gtrsim 5\times 10^4 L_{\odot}$ for $6-10 M_{\odot}$ progenitors.}
        \label{fig:DGL}
    \end{figure}

The main surprise in the invariance of PNLF bright-end cutoff is its consistency across galaxy types. Early-type galaxies typically have very low star formation rates compared to late-type galaxies. As a result, ellipticals in general have fainter central stars in PNe that formed from lower mass stars ($\lesssim 1M_{\odot}$), while spirals also contain more massive stars ($\gtrsim 6-10 M_{\odot}$) that evolve to become brighter post-AGB stars. The  PNe relevant to the bright-end cutoff of the PNLF will likely come from the most massive progenitor of each galaxy.  To have a consistent maximum [O III] luminosity thus requires a mechanism to counteract the difference in the luminosity of the brightest post-AGB stars in different galaxies.
Figure \ref{fig:PNLF} shows the observed bright-end cutoff value of $M_{5007} = -4.53\pm0.06$ translated to line luminosity from \citet{Ciardullo2012}. In our calculations, for models with $m_{\rm ej} = 0.1 M_{\odot}$ as an example, PNe hosting a $50,000 L_{\odot}$ post-AGB star (from a $\sim 6-10 M_{\odot}$ progenitor) and a $R_{\rm D/G} \simeq 2\%$ can produce similar $L^{\rm peak}_{\rm [O\,III]}$ compared to PNe hosting a $8000 L_{\odot}$ post-AGB star (from a $\sim 1-3 M_{\odot}$ progenitor) and a $R_{\rm D/G} \simeq 0.6\%$.   By setting $L^{\rm peak}_{\rm [O\,III]}$ to the observed PNLF cutoff values, we calculate the D/G ratio needed for a range of central stellar luminosities to reproduce the PNLF cutoff, as shown in Figure \ref{fig:DGL} for models with a few different ejecta masses. The required values for $R_{\rm D/G}$ shown here fall within a reasonable range of observed dust-to-gas mass ratios in PNe:  in \citet{SS1999}'s sample of 500 galactic PNe, 70\% have D/G between 0.1\% and 1\%, 15\% smaller than 0.1\%, and 15\% greater than 1\%.
In fact, \citet{SS1999} attribute the wide range of dust content observed in PNe populations to two possibilities: (1) PNe progenitor AGB stars carry different D/G mass ratios or (2) PNe with lower D/G mass ratios host a lower-mass star. The second scenario is consistent with our hypothesis for the origin of the observed constant maximum line luminosity in different galaxies.
We stress that according to our photoionization calculations, a trend like equation \ref{eq:LOIII} \& \ref{eq:taudust} and Figure \ref{fig:DGL} is the {\em only} way to produce a constant peak [O III] line luminosity with if there is significant variation in central star luminosity.   This is because the peak [O III] line luminosity depends strongly on only $L_\star$ and ${\rm R_{D/G}}$, with a weak dependence on other parameters, in particular the ejecta mass and proto-WD effective temperature (Figure \ref{fig:PNLF}). Note in particular that the dependence on ejecta mass is such that larger ejecta masses have somewhat larger $L^{\rm peak}_{\rm [O\,III]}/L_{\star}$, particularly at higher D/G mass ratios.   This trend implies that if more massive stars eject more gas in the PNe phase -- as suggested by the WD initial-mass final-mass relation \citep{Cummings2018} -- that more massive stars would have larger values  $L^{\rm peak}_{\rm [O\,III]}/L_{\star}$.   This trend is  {\em opposite} of that required to explain a consistent PNLF bright cutoff across galaxy types with very different stellar populations.   This again highlights the key role of a varying dust-to-gas mass ratio.

\section{The Contribution of WD Merger Remnants}
\label{sec:merger}

    \begin{figure*}
        \centering
        \includegraphics[width=0.48\textwidth]{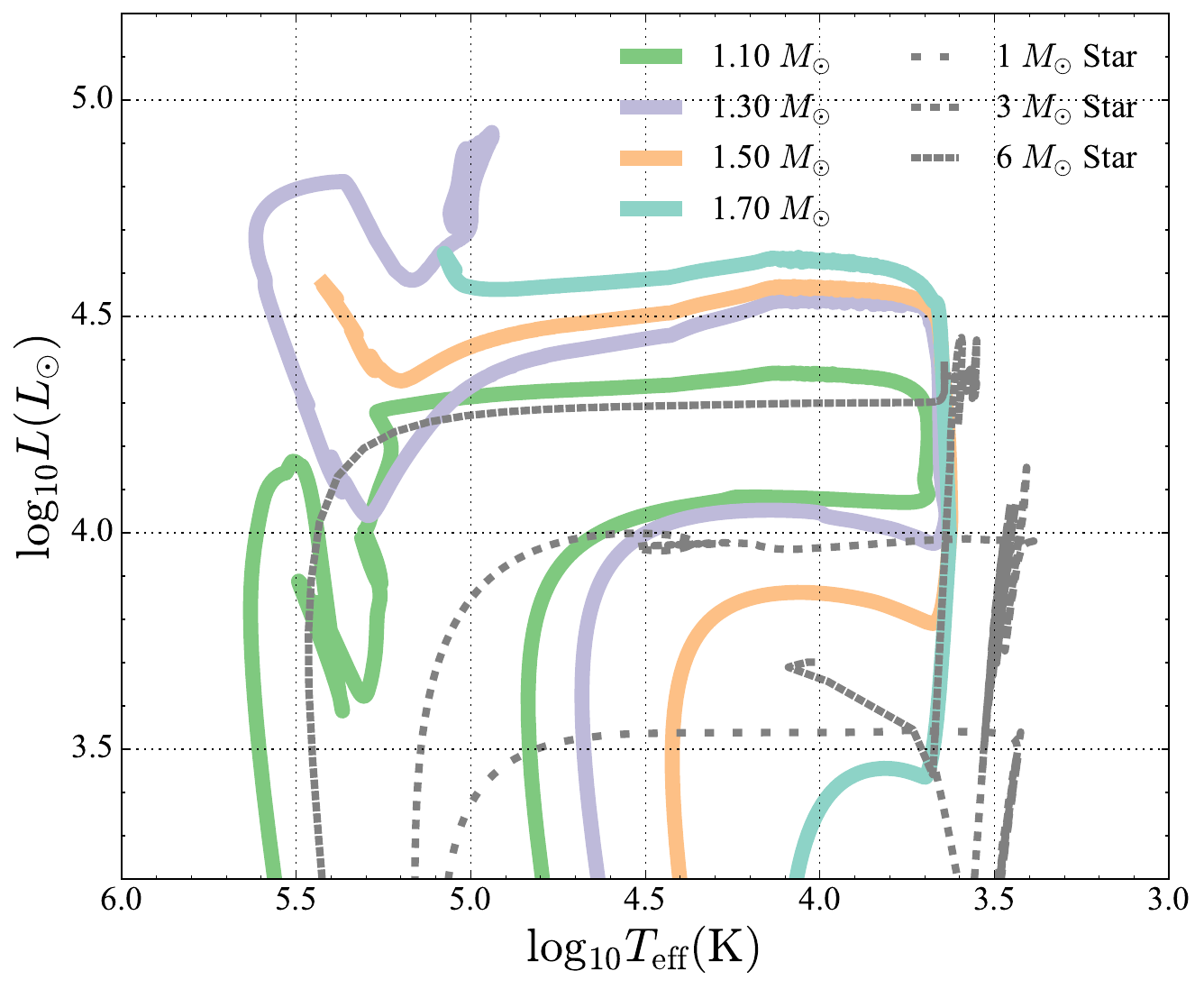}
        \includegraphics[width=0.5\textwidth]{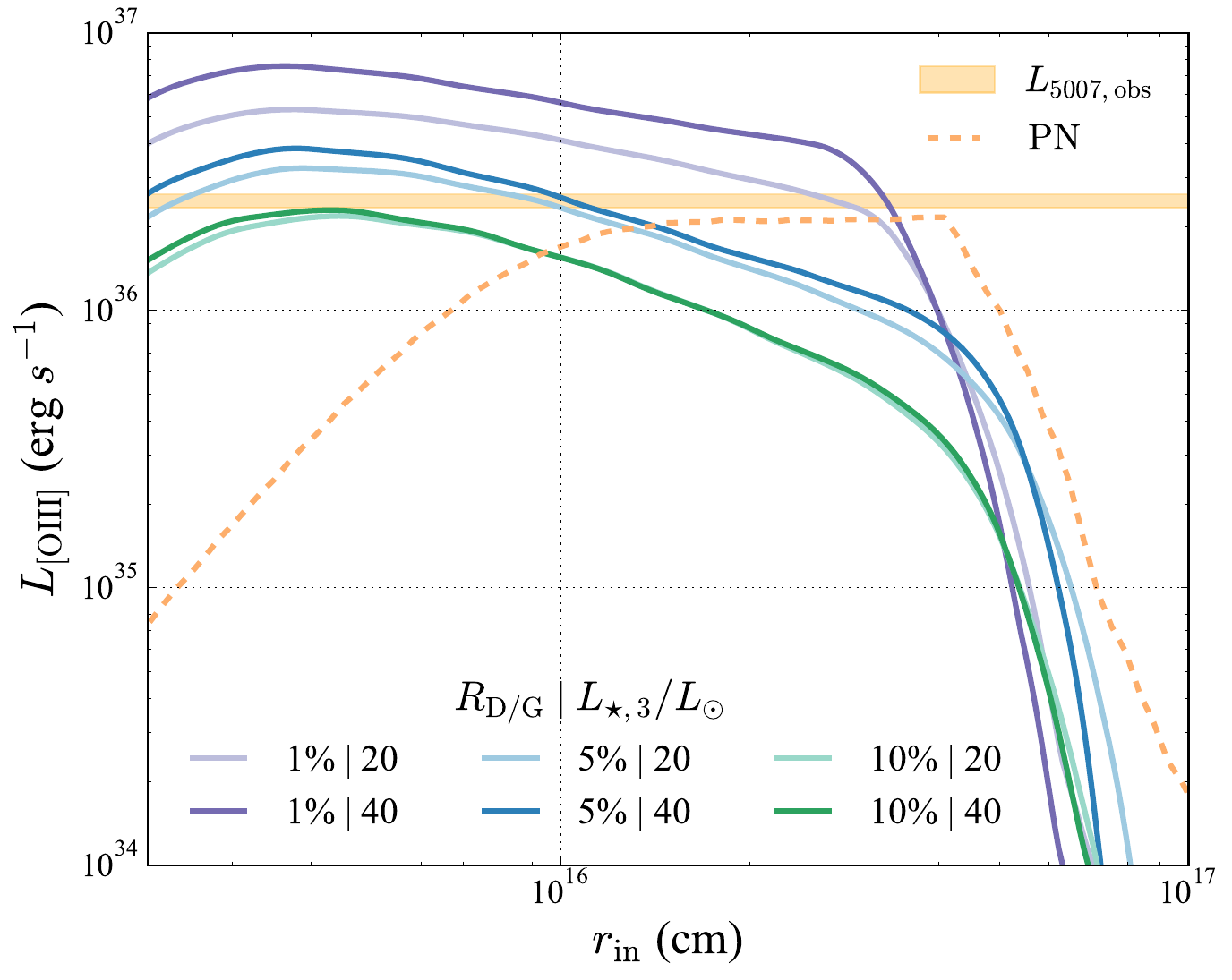}
        
        \caption{Carbon-oxygen white dwarf merger remnants appear similar to PNe and can produce bright [O III] lines. On the left panel, we show evolutionary tracks (solid lines) of CO WD merger remnants from \citet{Schwab2021} with a binary mass ratio of 0.9 and a range of total masses. For comparison, we overplot on the same panel evolutionary trajectories of typical post-main-sequence single stars (dotted lines) evolved with MESA from \citet{Farag2020}. On the right panel, we show WD merger remnants' $L_{\rm [O\,III]}$ from CLOUDY simulation as a function of cloud inner radii with a few different $R_{\rm D/G}$ and $L_{\star}$. This is analogous to the PNe version of the plot (Figure \ref{fig:radial}); we overplot a low-mass star PN model with $L_{\star,3} = 5 L_{\odot}$ and $R_{\rm D/G} = 1\%$ here (dashed line) for comparison. The WD merger remnant nebulae should be brighter than PNe from low mass stars in old stellar populations unless $R_{\rm D/G} \gtrsim 10 \%$.   The orange shaded region reflects the observed PNLF bright-end cutoff at $M_{5007} = -4.53\pm0.06$ translated to line luminosity from \citet{Ciardullo2012}.}
        \label{fig:merger}
    \end{figure*}

Although some double WD mergers very likely lead to a Type Ia supernova \citep{Webbink1984,Iben1984}, many probably do not and instead leave behind a long-lived remnant.  If the merger involves two CO WDs or a CO and ONeMg WD, the final fate of any surviving remnant depends on the total mass of the system.   If the total mass of the merger remnant (including the effects of post-merger mass loss) remains above the Chandrasekhar mass, the merger remnant ultimately collapses to form a neutron star \citep{Schwab2016}.    On the other hand, if the total mass is below the Chandrasekhar mass, the merger remnant ultimately becomes a cooling WD evolving down the WD cooling sequence \citep{Schwab2021}.   The left panel of Figure \ref{fig:merger} shows post-merger evolutionary models from \citet{Schwab2016,Schwab2021} of merger remnants with total masses from $1.1-1.7 M_\odot$.   For comparison, we also show single stellar models for $1, 3,$ and $6 \, M_\odot$ stars.  The total duration of the evolution shown for the merger remnants is $\sim 10^4$ yrs until the end of the nearly constant-luminosity phase.   

As the left panel of Figure \ref{fig:merger} shows surviving WD merger remnants evolve to become luminous hot proto-WDs, similar to the evolution of $6 M_\odot$ stars.   During their earlier giant phases (analogous to the AGB phase in single stellar evolution) it is very likely that the remnants lose significant mass.   \citet{Yao2023} showed that one of the most promising ways to detect WD merger remnants is via their photoionized nebulae analogous to PNe.   The right panel of Figure \ref{fig:merger} shows predictions of the [O III] luminosity from WD merger remnant nebulae for a few central stellar luminosities and D/G mass ratios.  The D/G mass ratios for these objects are particularly uncertain because of their unusual C and O-rich compositions.  The calculations in the right panel of Figure \ref{fig:merger} are from the models of  \citet{Yao2023} and are identical to the PNe calculations in Figure \ref{fig:radial} except they assume higher $L_\star$ and nebula compositions appropriate to CO WD ejecta.   These results show that WD merger remnants can produce [O III] luminosities at least as large as standard PNe.   In an old stellar population, the WD merger remnant nebulae would plausibly be brighter than the PNe from $1 M_\odot$ stars, unless the D/G ratio is $R_{\rm D/G} \gtrsim 10 \%$ 

How many WD merger remnant nebulae are there compared to PNe from single stellar evolution? For a Kroupa IMF there are approximately $2 \times 10^{10}$ sun-like stars for a system with a total stellar mass of $M_\star = 10^{11} M_\odot$. Since the lifetime of such stars is $\simeq 10^{10}$ yrs the formation rate of PNe progenitors from 1 $M_\odot$ stars is
\begin{equation}
\dot N_{1 M_\odot} \simeq 2 \, {\rm yr^{-1}} \, \left(\frac{M_\star}{10^{11} M_\odot}\right)
\label{eq:Ndot1}
\end{equation}
For comparison, if we scale the surviving merger remnant rate to the Type Ia rate (as a proxy for the WD merger rate, which is harder to directly constrain), we estimate
\begin{equation}
\dot N_{\rm merger} \simeq 4 \times 10^{-2} \, {\rm yr^{-1}} \, \left(\frac{M_\star}{10^{11} M_\odot}\right) \left(\frac{\dot N_{\rm merger}}{\dot N_{\rm Ia}}\right)
\label{eq:Ndotmerger}
\end{equation}
where we have used the Type Ia rate associated with an old stellar population from \citet{Scannapieco2005}.   Note that since the duration of the hot luminous phase in Figure \ref{fig:merger} is $\sim 10^4$ yr, a rough estimate of the total number of WD merger remnant nebulae in a stellar population is $\sim 400 \, (M_\star/10^{11} M_\odot)$. Bright remnants that might contribute to the bright-end cutoff are likely rarer by perhaps an order of magnitude. In more detail, the number of remnants will vary with [O III] luminosity in a way that is sensitive to the speed of winds produced during the giant phase.

A comparison of equation \ref{eq:Ndot1} and \ref{eq:Ndotmerger} shows that WD merger remnant nebulae have a formation rate $\sim 50$ times smaller than PNe from 1 $M_\odot$ stars. Since the merger remnants are more luminous, however, their resulting nebulae may be somewhat brighter in  [O III] (Figure \ref{fig:merger}) and thus could contribute preferentially to the bright end of the [O III] luminosity function in old stellar populations.   The population of WD merger remnants should be much more similar across galaxy types than PNe from single stellar evolution given that the WD merger rate is expected to be more consistent across galaxy types than the star formation rate.   The contribution of WD merger remnant nebulae to the [O III] PNLF could be most easily quantified via the detection of or limits on the H$\alpha$/[O III] line ratio.

A second comparison that is valuable is between the rate of formation of WD merger remnants and the formation rate of WDs from $\sim 6 M_\odot$ stars in a stellar population with ongoing star formation.   These two channels produce proto-WDs of similar luminosity and effective temperature and may produce PNe with comparable [O III] line luminosities (Fig. \ref{fig:merger}).  For a Kroupa IMF the formation rate of 6 $M_\odot$ stars is given by
\begin{equation}
\dot N_{6 M_\odot}  \simeq \, 0.02 \, {\rm yr^{-1}} \, \left(\frac{\dot M_\star}{1 \, M_\odot \, {\rm yr^{-1}}}\right).
\label{eq:6Msun}
\end{equation}
A comparison of equations \ref{eq:6Msun} and \ref{eq:Ndotmerger} shows that for star-forming galaxies with star-formation rates of a few solar masses per year, the formation rate of WDs from massive stars may be similar to the formation rate of double WD merger remnants in massive old stellar populations.   Thus, at least at the order of magnitude level, this coincidence could contribute to the similar bright-end cut cutoff in the PNLF across stellar populations.

\section{Discussion}\label{sec:discussion}
In this paper, we show that photoionization modeling of PNe suggests a solution to the invariance of the bright-end cutoff of the [O III] PNLF across galaxy types. We suggest that higher dust-to-gas mass ratios at higher WD luminosities (higher progenitor stellar masses) effectively conspires to produce a roughly constant maximum [O III] luminosity across galaxy types. By contrast, a larger ejecta mass at higher progenitor stellar mass, as suggested by the WD initial mass final mass relation, produces only a modest change in peak PNe [O III] luminosities, and in a direction inconsistent with an invariant bright end cutoff (higher ejecta masses from higher mass more luminous progenitors would produce a larger not smaller ratio of $L^{\rm peak}_{\rm [O III]}$; see Fig. \ref{fig:PNLF}).    Somewhat more speculatively, we also suggest that photoionized nebulae from the surviving remnants of double WD mergers may contribute to the bright end of the [O III] PNLF.   The population of WD merger remnant nebulae is likely to be relatively similar across galaxy types (analogous to how the Type Ia rate is much more similar across galaxy types than the core-collapse supernova rate).  

A second problem related to the consistent PNLF bright-end cutoff is the observed maximum $M_{5007} \simeq -4.5 \, \rm mag$ in early-type galaxies.  This [O III] luminosity is in some tension with standard stellar evolutionary models and the initial-final-mass relation (e.g., \citealt{Davis2018,Kwitter2022}). For example, our CLOUDY simulations show that a minimum central stellar luminosity of $\sim 5000 L_{\odot}$ with a negligible amount of dust ($\sim 0.1\%$) is just capable of producing the observed cutoff [O III] luminosity in early-type galaxies.  Solar mass stars are expected to produce proto-WDs somewhat less luminous than this. Newer post-AGB evolution models \citep[e.g.][]{Miller2016} may alleviate this tension. In addition, we note again that surviving WD merger remnants could be important at the luminous end of the [O III] luminosity function in early-type galaxies, potentially explaining the population of PNe more luminous in [O III] than expected in standard single-star models.  This possibility can be tested using deep H Balmer-line observations of [O III]-luminous nebulae in early-type galaxies. Put another way, observational constraints on the number of photoionized nebulae with high [O III]/H-$\alpha$ in massive galaxies could directly constrain the uncertain fraction of WD mergers that leave behind long-lived remnants. The [O III] and H$\alpha$ lines for PNe should be detectable above the stellar continuum in lower surface brightness nearby galaxies and galaxy clusters such as in the bulge of M31 and the outskirts of M87 (see \citealt{Yao2023} for a more detailed discussion of this point). In contrast, WD merger remnants would appear [O III] bright with no detectable-H Balmer emission from a photoionized nebula surrounding a hot central object.

The possible variation in the dust-to-gas mass ratio with stellar luminosity in PNe can also be tested. The dust-to-gas ratio, $R_{\rm D/G} \propto A_{\rm V}/N_{\rm H}$, can be obtained by separately measuring the extinction and line-of-sight hydrogen column density. $A_{\rm V}$ can be measured from the ratio of H$_{\alpha}$/H$_{\beta}$ lines, and $N_{\rm H}$ can be estimated, given a small neutral fraction, using the H$\beta$ flux, plus electron density and temperature from line diagnostics \citep[see a detailed outline in][]{Walsh2016}. Performing a survey of PNe dust-to-gas mass ratios as a function of central stellar luminosity will indicate if our proposed correlation indeed holds true. While these quantities can be challenging to measure in extragalactic PNe populations due to the high stellar continuum, the Milky Way contains a large sample of PNe suitable for this analysis.

\section*{Acknowledgements}
We thank Bruce Draine for helpful conversations and Robin Ciardullo for very helpful comments on an initial draft of the paper. The referee's thoughtful comments significantly improved the manuscript. EQ was supported in part by a Simons Investigator award from the Simons Foundation.   This research benefited from interactions at workshops funded by the Gordon and Betty Moore Foundation through Grant GBMF5076.


\bibliography{sample631}{}
\bibliographystyle{aasjournal}



\end{document}